\documentclass[showpacs,showkeys,aps,prc,preprintnumbers,amsmath,amssymb,superscriptaddress]{revtex4}

\usepackage{graphicx}
\usepackage{dcolumn}
\usepackage{bm}
\usepackage{subfig}
\usepackage[usenames]{color}
\usepackage{epsfig,pstricks,dcolumn}

\begin{document}

\date{\today}

\title{ Cluster virial expansion for nuclear matter within 
a quasiparticle statistical approach}

\author{G. R\"opke}
\affiliation{Universit\"at Rostock, Institut f\"ur Physik, 18051 Rostock, 
Germany}
\author{N.-U. Bastian}
\affiliation{Universit\"at Rostock, Institut f\"ur Physik, 18051 Rostock, 
Germany}
\author{D. Blaschke}
\affiliation{Instytut Fizyki Teoretycznej, Uniwersytet Wroc\l{}awski, 
pl. M. Borna 9, 50-204 Wroc\l{}aw, Poland}
\affiliation{Bogoliubov Laboratory for Theoretical Physics, JINR Dubna, 
Joliot-Curie str. 6, 141980 Dubna, Russia}
\author{T. Kl\"ahn}
\affiliation{Instytut Fizyki Teoretycznej, Uniwersytet Wroc\l{}awski, 
pl. M. Borna 9, 50-204 Wroc\l{}aw, Poland}
\author{S. Typel}
\affiliation{GSI Helmholtzzentrum f\"ur Schwerionenforschung GmbH, Theorie, 
Planckstra{\ss}e 1, D-64291 Darmstadt, Germany}
\author{H.H. Wolter}
\affiliation{Fakult\"at f\"ur Physik, Universit\"at M\"unchen, 
Am Coulombwall 1, D-85748 Garching, Germany}

\begin{abstract}
 Correlations in interacting many-particle systems can lead to 
the formation of clusters, in particular bound states and resonances.
Systematic quantum statistical approaches allow to combine the nuclear 
statistical equilibrium description (law of mass action) with mean-field concepts.
A chemical picture, which treats the clusters as distinct entities,
serves as an intuitive concept to treat the low-density 
limit. Within a generalized Beth-Uhlenbeck approach, the quasiparticle virial 
expansion is extended to include arbitrary clusters, where special 
attention must be paid to avoid inconsistencies such as double counting.
Correlations are suppressed with increasing density due to Pauli blocking.
The contribution of the continuum to the virial coefficients can be reduced by 
considering clusters explicitly and introducing quasiparticle energies.
The cluster-virial expansion for nuclear matter joins known benchmarks 
at low densities with those near saturation density.

\end{abstract}

\maketitle

\section{Introduction} 
\label{sec:introduction}
Recently considerable efforts have been made to elaborate the nuclear matter 
equation of state (EoS) in a wide range of baryon density $n$, temperature $T$,
and  proton fraction $Y_p$ (or neutron-proton asymmetry $\delta = 1-2Y_{p}$)
\cite{EoS,Klahn:2006ir,LS,unsere,Shen1998,Hempel2010,VT}. 
We consider in this work warm dilute matter 
($T \le 20$ MeV, $n\le n_0$ with $n_0 \approx 0.16$ fm$^{-3}$ being the nuclear
saturation density). 
In this
region the simple model of an ideal Fermi gas of particles has to be improved by including correlations
between the particles. 
It has to be emphasized that the notion of correlations depends on the
reference state.
For an ideal Fermi gas, the many-body state is given by a Slater
determinant of 
single-particle plane waves without any correlations beyond those
originating from quantum statistics. 
Explicit correlations become less significant near
the saturation density
because correlations are increasingly blocked
with increasing density and Fermi energy
and the many-body state can be
considered as a system of uncorrelated quasiparticles.
At low densities an explicit treatment of correlations is essential. 
In particular, clusters can dominate the composition of matter at finite density at
 low 
temperatures. 
Different approximations have been considered to treat the formation of 
clusters in the low-density limit of the nucleon gas, such as the nuclear 
statistical equilibrium (NSE)  approach \cite{NSE,Hempel2010} or virial expansions 
\cite{Beth:1936zz,Beth:1937zz,Sedrakian:1997zd,HS}. 
Here we want to show that a
consistent treatment of correlations (bound as well as scattering states) 
can be given within a quantum statistical approach. 
 
The main question is to match quasiparticle concepts with cluster
concepts, as e.g.\ embodied in the law of
mass action at low densities. 
This has been discussed for the two-nucleon problem within a generalized 
Beth-Uhlenbeck approach \cite{SRS} which was developed originally for electron-hole 
plasmas \cite{Zimmermann:1985ji}.
A general expression for the second virial coefficient was obtained that 
contains in addition to the contribution of bound states also those  
of scattering states. 
As an important result, it has been demonstrated that the contribution of the 
continuum of scattering states is modified when quasiparticles are introduced.
 Larger clusters consisting of $A$ nucleons were 
 introduced via a cluster 
decomposition of the nucleon self energy \cite{RMS}. A quasiparticle 
treatment of nuclei was considered via an in-medium Schr\"odinger equation,
where, in addition to the single-nucleon self-energy shift, also the 
Pauli blocking was considered. 
In-medium quasiparticle energies for the  ground states of light nuclei $A\le 4$ have been 
given recently \cite{R2011} as function of temperature $T$, baryon density $n$,
proton fraction $Y_p$, and the center-of-mass momentum $P$. 
The solution of the in-medium few-body problem would also give excited states 
as well as the scattering states.

Quantum statistical approaches are based on perturbation expansions. 
Within the Green function method, Feynman diagrams are introduced, and 
partial summations are performed. 
We focus on a special prescription to select relevant contributions of the 
perturbation expansion that describe correctly the formation of clusters in 
the low-density limit.
The concept to include bound states on the same footing as new particle species
 that can react is 
called the {\em chemical picture}. 
 On the other hand, a fundamental quantum statistical approach of interacting ”elementary”
particles is the {\em physical picture}. From this the chemical picture is obtained if 
in addition to diagrams with single-particle (quasi-particle) propagators also 
diagrams are considered where the single-particle propagators are replaced by  
ladder-type diagrams that describe the propagation of the $A$ particle cluster. 
If only the bound state contributions are included, one obtains the NSE in the low-density
limit.

If scattering contributions are included in the two-body channel, one obtains the virial expansion.
A generalization of the virial expansion is also possible within the chemical 
picture if in any channel of the cluster-cluster interaction not only the 
formation of bound states, but also the contribution of scattering states to 
the thermodynamical properties of the system is considered. 
Indeed, there are regions in the density-temperature plane where clusters are dominant 
and should be considered as new constituent particles within the chemical 
picture. 
In this case, a cluster-virial expansion
should be performed to include the effects of 
the continuum. 
Such an approach is evident from the empirical point of view, but not easily 
derived from first principles. 
The use of a cluster mean-field (CMF), as well as nearly bound states
from the continuum correlations can lead to double counting, 
and these contributions have to be extracted from the
continuum contribution.

Recently, such a cluster-virial expansion has been applied to the 
proton-neutron-$\alpha$ system \cite{HS} and was extended
to include additional light clusters \cite{OCo07}. 
Bound states, such as the deuteron or resonances like $^8$Be, 
were treated differently, 
sometimes as part of the generalized second virial coefficient, 
sometimes as new constituent particles. A systematic
derivation of a cluster virial expansion that includes as a limit also the NSE 
has to be performed on the basis of a fundamental quantum statistical 
approach.  
Contributions occurring in higher-order virial coefficients that correspond to 
bound states or resonances have to be separated since they are explicitly 
accounted for in the NSE including all nuclei in ground and excited states.

The chemical picture can also be used to derive a cluster-mean field 
approximation that treats the mean-field effects of a correlated medium 
\cite{cmf}. 
It is expected that the incorporation of mean-field effect into the cluster 
quasiparticle states will also change the contribution of the scattering 
states in the cluster-virial expansion. 
 This can be understood
from a fundamental point of view by considering the spectral function in the respective A-nucleon channel. 
Cluster-quasiparticles should be 
introduced representing
the peaks of the $A$-nucleon spectral function. 
Then the explicit treatment of the contribution of these peaks 
as quasiparticles will account for a significant part of the total 
contribution. 
This is clearly seen in the low-density region where a law of mass action
can be introduced. 
The occurrence of bound states in the EoS according to the
law of mass action
is a signature that significant 
contributions have to be extracted from the spectral function of the 
elementary nucleons. 
The ordinary quasiparticle picture where the nucleonic spectral function is 
assumed to be sharply ($\delta$-like) peaked is no longer justified.

The inclusion of scattering states is of importance for the equation of state 
as well as for further properties of nuclear systems. 
Recently, the composition of low-density nuclear matter 
was investigated \cite{unsere} and it was seen in the quantum
statistical approach that the contribution of scattering states led 
to a reduction of the deuteron mass 
fraction. 
This has to be taken into account in particular at high temperatures. 
We will give some details here. 
The main result is that the contribution of the continuum can be reduced if 
mean-field and cluster contributions are already extracted.

Some results of the  chemical picture are given in Sec.~\ref{Sec:2}, 
however a comprehensive treatment cannot be given here. 
We discuss the different effects and propose some approximations. 
The main aim of this paper is to obtain a  more systematic treatment in particular of the 
cluster-virial expansion and to avoid inconsistencies such as double counting.
As example, the second virial coefficient is considered in Sec.~\ref{Sec:3} 
that gives the leading contribution in the low density limit.
In particular, we discuss the ambiguities connected with the 
introduction of the bound state contribution of the virial coefficient
and the consistent introduction of the quasiparticle picture.
The relation to the generating functional approach that allows for 
a systematic quantum statistical approach to thermodynamic properties 
is outlined in Sec.~\ref{Sec:4}. 
Conclusions are drawn in Sec.~\ref{Sec:5}, and 
details are given in the Appendices.

\section{The chemical picture and correlations}
\label{Sec:2}

We explain first some general results for many-particle systems.
In the {\em chemical picture}, we start with a mixture of different 
constituents that can be the elementary particles (atoms, here nucleons) as 
well as the bound states (molecules, here nuclei). The main issue is to design
approximations where ``elementary'' particles and ``composite''
particles are
treated on the same footing. This concept has to be incorporated into a 
quantum statistical approach using Green functions techniques. 
The cluster Green function approach is briefly summarized in
App.~\ref{App:cese}.

\subsection{Nuclear statistical equilibrium}
\label{subsec:NSE}

At low densities we can neglect the interaction with exception of collisions 
where the constituents come close together resulting in reactions 
that establish the chemical equilibrium. 
These reactions include excitations and ionization, e.g., in a plasma. 
The same considerations apply also for nuclear systems where various nuclei 
occur that can react.
Thus, we have as approximation an ideal mixture of different components.

As a result, the law of mass action is 
found, with the total particle (baryon) number density
(note that astrophysical
$\beta$-equilibrium is not considered here)
\begin{equation}
\label{eq:idgas}
 n(T,\mu_{p},\mu_{n}) 
 =  \sum_{A,Z,\nu} \frac{A}{\Omega} \sum_{\vec{P}} 
 f_{A}(E_{A,Z,\nu}^{(0)}(\vec{P}),\mu_{A,Z})
\end{equation}
where $A,Z$ denote mass and charge number of a nucleus, respectively, and 
$\nu$ indicates the internal quantum state of the nucleus.
$\vec{P}$ is the center of mass momentum,
$\Omega$ is the volume of the system, and
\begin{equation}
f_{A}(E,\mu)=\frac{1}{\exp [(E - \mu)/T]- (-1)^{A}}
\label{vert}
\end{equation}
is the Bose or Fermi distribution function for even or odd $A$,
respectively. Note that at low temperatures Bose-Einstein condensation
may occur, which is neglected here.
We determine the chemical potentials $\mu_{A,Z}
= Z\mu_{p}+(A-Z)\mu_{n}$ only with respect to the 
kinetic energy of species $\{A,Z,\nu\}$, the binding energy  
$B_{A,Z,\nu}$ 
is considered explicitly.

In the low-density limit, the energies
\begin{equation}
 E^{(0)}_{A,Z,\nu}(\vec{P})= -B_{A,Z,\nu}+ P^2/(2m_{A,Z,\nu})
\end{equation}
are given by the binding energies 
$B_{A,Z,\nu}$ of the isolated nuclei in vacuum
with masses $m_{A,Z,\nu}=Zm_{p}+(A-Z)m_{n}-B_{A,Z,\nu}$, denoted by the index (0).
In the NSE the summation over $\nu$ concerns only bound states, the 
contribution of the continuum is neglected. 
Neglecting medium corrections, the ordinary law of mass
action will increase the 
mass fraction of bound states like $d$ and even more the $\alpha $ particle 
with respect to the free
proton and neutron fractions when the total nucleon number
density $n$ increases. This behavior contradicts 
the expectation that at high 
densities a nucleonic quasiparticle picture is appropriate
without the explicit occurrence of clusters.

For a more fundamental approach that is not limited to the low-density limit, 
one should use the {\em physical picture} where some of the constituents are 
elementary, while others are composite particles. 
The composite particles are obtained as bound states of the elementary 
particles. 
The same interaction potential, that leads to the formation
of bound states in the solution of the many-body Schr\"odinger equation,
determines also the 
interaction between the constituents of the system. 
(Note that it is common to formulate statistical physics with 'elementary' 
particles that on their part are composed of more elementary particles.) 
The simple law of mass action with clusters
where the interaction between the components is 
neglected (with exception of reactive collisions) works well in the low-density
region and low temperatures, in contrast to a picture of an 
ideal mixture of the elementary 
particles such as the ideal Fermi gas of protons or neutrons.

\subsection{ Cluster-virial approach}
\label{subsec:cva}

We discuss three issues that are of relevance to improve the NSE. 
The first one refers to the summation over excited states that, formally, can 
also be considered as new, independent constituents.
Contributions of the continuum are neglected within the chemical picture, and 
only bound clusters are considered as new `components'. 
The limits of this simple chemical picture are evident for 
sharp resonances above the continuum threshold.
Long living states should be included in a more general treatment if the life 
time (inverse of the width) is sufficiently long. 
We will consider this problem below within the physical picture using a quantum 
statistical approach. To obtain the complete second virial coefficient, however,
the contribution of scattering states has to be included according to
the Beth-Uhlenbeck approach. 

The second issue  refers to the treatment of the interaction between the 
different components. 
The chemical picture takes interactions into account only to 
establish chemical equilibrium.
Effects of scattering correlations on the thermodynamical
quantities are neglected. 
To obtain the full second virial coefficient, the contribution of scattering 
states has to be included. 
Nevertheless, the chemical picture describes well the low-density, 
low-temperature region. 
It can be improved taking so-called excess terms in the chemical potential and 
other thermodynamic variables into account. 
As an example, a simple way to include non-ideal effects is the excluded 
volume concept  \cite{Hempel2010}.
More systematic approaches consider scattering phase shifts due to
the 
interaction between the constituents, as considered in 
the 
Beth-Uhlenbeck formula. 
Then, the total baryon number density
\begin{equation}
\label{virial}
 n(T,\mu_{p},\mu_{n})
 = n_{1}(T,\mu_{p},\mu_{n})
 + n_{2}(T,\mu_{p},\mu_{n})
 + n_{3}(T,\mu_{p},\mu_{n}) + \dots
\end{equation}
contains a contribution $n_{1}$ from the individual constituents, i.e.\
nucleons and bound states of nuclei. This term is
identical to the NSE result Eq.(\ref{eq:idgas}). The contributions
$n_{2}$, $n_{3}$, \dots  account for the two-body, three-body, \dots correlation effects in the
continuum. In the original Beth-Uhlenbeck formulation there are
contributions from bound and scattering two-body correlations
appearing in $n_{2}$. However, in order not to count contributions
twice as original
constituents in $n_{1}$ and as bound states in $n_{2}$, 
the latter term should contain only the scattering part. It can be
expressed through
integrals with scattering phase shifts $\delta_{c}$ in all channels
$c$ for the scattering of nuclei $\{A,Z,\nu\}$ and
$\{A^{\prime},Z^{\prime},\nu^{\prime}\}$ . Thus it is given by
\begin{equation}
\label{eq:n2_vir}
 n_{2}(T,\mu_{p},\mu_{n})  =  \sum_{A,Z,\nu}
 \sum_{A^{\prime},Z^{\prime},\nu^{\prime}} \frac{A+A^{\prime}}{\Omega} \sum_{\vec{P}}
 \sum_{c} g_{c}\frac{1+\delta_{A,Z,\nu ;
     A^{\prime},Z^{\prime},\nu^{\prime}}}{2\pi}  \int_{0}^{\infty} dE \:
 f_{A+A^{\prime}}\left(E_{c}^{(0)}(\vec{P})+E,\mu_{A,Z}+\mu_{A^{\prime},Z^{\prime}}\right) 
 \frac{d\delta_{c}}{dE}
\end{equation}
with energies 
\begin{equation}
 E_{c}^{(0)}(\vec{P}) = -B_{A,Z,\nu}-B_{A^{\prime},Z^{\prime},\nu^{\prime}}
+\frac{P^{2}}{2(m_{A,Z,\nu}+m_{A^{\prime},Z^{\prime},\nu^{\prime}})}
\end{equation}
and degeneracy factors $g_{c}$.

The third issue is related to
medium effects. The energies $E^{(0)}_{A,Z,\nu}(\vec{P})$ 
and $E^{(0)}_{c}(\vec{P})$ in the NSE and the
virial expansion, respectively, will be replaced by quasiparticle energies 
$E_{A,Z,\nu}(\vec{P};T,\mu_p,\mu_n)$ of the clusters leading to the
generalized cluster Beth-Uhlenbeck approach that is discussed in the
next subsection.
These quasiparticle energies
depend on density and temperature in many-particle
systems of finite density.
Medium effects become operative if the density exceeds about  $10^{-4}$ fm$^{-3}$. 
In particular, Pauli blocking will dissolve the clusters at densities about  
$10^{-2}$ fm$^{-3}$ \cite{R2011}.

The quasiparticle approximation has been used in several recent approaches to 
calculate the properties of nuclear matter \cite{Klahn:2006ir,LS,Shen1998}. 
In particular, it is very effective near saturation density but fails in the 
low-density region where clusters eventually become of importance. 
As a first step, we discuss the inclusion of two-particle correlations and the 
generalized Beth-Uhlenbeck formula, i.e. the $n_2$ term in
Eq.(\ref{virial}), in the following Section. 
It reproduces the correct second virial coefficient but gives also the correct 
high-density limit.
This allows to reproduce the NSE in the low-density limit but also the 
transition to the correct behavior in the high-density region. 

It will be shown, that both effects, the consideration of the interaction between clusters, 
i.e.\ of cluster mean-field effects, as well as the taking into account 
of excited states and resonances, will reduce the contribution of the
continuum to the equation of state, so that appropriate approximations can be obtained.

\subsection{Generalized cluster Beth-Uhlenbeck approach}
\label{genBU}

Now we discuss the inclusion of medium effects. 
The effective degrees of freedom are now quasiparticles with self-energies
that replace the original constituents.
The full evaluation of the imaginary part of the single-nucleon
  self-energies
${\rm Im}\,\, \Sigma_1$ in the so-called $T_2\,G_1$ approximation, see Ref.\ \cite{SRS},
leads to a generalized  Beth-Uhlenbeck equation and  
corresponding EoS. We can extend this approach, originally formulated
with only nucleons as basic constituents, heuristically to consider nucleon and
cluster quasiparticles. A systematic treatment based on a generating
functional approach is sketched in Section \ref{Sec:4}.

Considering two-body correlations of nucleons and clusters at
  most, the total baryon number density
\begin{equation}
\label{qpvirial}
 n(T,\mu_{p},\mu_{n})
 = n_{1}^{\rm qu}(T,\mu_{p},\mu_{n})
 + n_{2}^{\rm qu}(T,\mu_{p},\mu_{n})
\end{equation}
receives contributions from single quasiparticles ($n,\,\,p,\,\,d,\,\,t,\,\,^3$He, $^4$He, etc.) and correlated
two-quasiparticle continuum states. The one-body term
\begin{equation}
\label{genclBU}
 n_{1}^{\rm qu}(T,\mu_{p},\mu_{n}) 
 =  \sum_{A,Z,\nu} \frac{A}{\Omega} \sum_{\genfrac{}{}{0pt}{}{\vec{P}}{P>P_{\rm Mott}}}
 f_{A}\big(E_{A,Z,\nu}(\vec{P};T,\mu_{p},\mu_{n}),\mu_{A,Z,\nu}\big)
\end{equation}
resembles the corresponding contribution (\ref{eq:idgas}) 
in the NSE and $n_{1}(T,\mu_{p},\mu_{n})$ in the conventional
cluster-virial approach discussed in
subsections \ref{subsec:NSE} and \ref{subsec:cva}. However, there are
distinct differences. The quasiparticle energies
\begin{equation}
 E_{A,Z,\nu}(\vec{P};T,\mu_{p},\mu_{n})=E_{A,Z,\nu}^{(0)}(\vec{P})
+\Delta E_{A,Z,\nu}^{\rm SE}(\vec{P};T,\mu_{p},\mu_{n})
\end{equation}
depend not only on the 
c.m.\ momentum $\vec{P}$ of the nucleon or cluster 
with respect to the medium but also on
the temperature and chemical potentials (or equivalently densities)
through the medium dependent self-energy shift 
$\Delta E_{A,Z,\nu}^{\rm SE}(\vec{P};T,\mu_{p},\mu_{n})$.
See App.~\ref{App:CMF} and Ref. \cite{R2011} for a derivation of these shifts.
In addition, bound states of clusters exist only for c.m.\ momenta $P$
that are larger than the (quasiparticle and medium dependent) Mott momentum 
$P_{\rm  Mott}$. This is mainly a 
consequence of the Pauli principle that suppresses the formation of
clusters due the population of low-momentum states by nucleons of the 
medium.
The Mott momentum $P_{\rm Mott}(T,n_n,n_p)$ indicates the critical momentum where the 
bound state merges with the continuum of scattering states. 
The Mott momentum becomes larger than zero if the density exceeds a critical 
value, i.e.\ above this density bound states can exist only for 
$P >P_{\rm Mott}(T,n_n,n_p)$ \cite{RMS}.
Obviously, $P_{\rm Mott} = 0$ for nucleons.

The two-quasiparticle scattering contribution
\begin{eqnarray}
\label{eq:n2qu}
 n_{2}^{\rm qu}(T,\mu_{p},\mu_{n}) &=& \sum_{A,Z,\nu}
 \sum_{A^{\prime},Z^{\prime},\nu^{\prime}} \frac{A+A^{\prime}}{\Omega} \sum_{\vec{P}}
 \sum_{c} g_{c}\frac{1+\delta_{A,Z,\nu ;
     A^{\prime},Z^{\prime},\nu^{\prime}}}{2\pi} 
 \\ \nonumber & & \times \int_{0}^{\infty} dE \:
 f_{A+A^{\prime}}\left(E_{c}(\vec{P};T,\mu_{p},\mu_{n})+E,\mu_{A,Z}+\mu_{A^{\prime},Z^{\prime}}\right) \:
 2  \sin^2\left( \delta_{c} \right) 
 \frac{d\delta_{c}}{dE}
\end{eqnarray}
is also modified in comparison to the virial result (\ref{eq:n2_vir})
since the energy
\begin{equation}
 E_{c}(\vec{P};T,\mu_{p},\mu_{n}) = 
 -B_{A,Z,\nu}-B_{A^{\prime},Z^{\prime},\nu^{\prime}}
 +\frac{P^{2}}{2(m_{A,Z,\nu}+m_{A^{\prime},Z^{\prime},\nu^{\prime}})} 
 + \Delta E_{c}^{\rm SE}(\vec{P};T,\mu_{p},\mu_{n})
\end{equation}
contains the medium-dependent shift $\Delta
E_{c}^{\rm SE}(\vec{P};T,\mu_{p},\mu_{n})$
of the continuum edge as determined by the
the self-energy shifts of the free constituents.
Moreover, there is an additional 
$ 2 \left[ \sin\left( \delta_{c} \right) \right]^{2}$
factor that reduces the two-body scattering contribution because a part of the
two-body correlation effect is shifted to the self-energies of the quasiparticles.

Note that the two-particle contribution (\ref{eq:n2qu}) describes binary scattering 
processes $A,\,\,A'$. A possible bound state contribution in that channel is 
excluded as contribution to the two-quasiparticle scattering contribution (\ref{eq:n2qu}).
The contribution of a possible $A+A'$ bound state is already taken into account in
the one-body term (\ref{genclBU}). Thus, double counting is avoided.

\mbox{}

\section{Comparison of approaches at low densities}
\label{Sec:3}

In the low-density limit, medium effects can be neglected. 
Only neutrons, protons and deuterons are the relevant constituents.
The standard Beth-Uhlenbeck formula for the 
second virial coefficient and the virial expansion
\cite{Beth:1936zz,Beth:1937zz}
are exact results. 
They are derived by expanding the thermodynamic functions with respect to the 
fugacities. 
It is instructive to show explicitly the equivalence with the generalized 
cluster Beth-Uhlenbeck approach that introduces the concept of quasiparticles 
to account for part of the correlations by introducing self-energies.

\mbox{}

\subsection{Cluster virial approach}

Let us first consider the cluster virial method applying a fugacity expansion
up to second order
in the chemical potentials of protons and neutrons. With the continuum
approximation $(1/\Omega) \sum_{\vec{P}} \to \int d^{3}P/(2\pi)^{3}$ the
one-body contribution assumes the form 
\begin{equation}
\label{eq:n1}
 n_{1} = n_{1}^{(p)} + n_{1}^{(n)} + n_{1}^{(d)}
\end{equation}
with the single nucleon contributions ($i=p,n$, including degeneracy effects)
\begin{equation}
 n_{1}^{(i)} = \frac{2}{\Lambda_{i}^{3}} \left[ \exp \left(\frac{\mu_{i}}{T}\right)
 - 2^{-3/2} \exp \left(\frac{2\mu_{i}}{T}\right) \right]
\end{equation}
and the single deuteron contribution
\begin{equation}
\label{nd1}
 n_{1}^{(d)} = 
 \frac{3}{\Lambda_{d}^{3}} \exp
 \left(\frac{\mu_{p}+\mu_{n}+B_{2,1}}{T}\right) \: .
\end{equation}
$\Lambda_{i} = \sqrt{2\pi/(m_{i}T)}$ are the thermal wavelengths
of the particles $i=p,n,d$ and $B_{2,1}>0$ is the binding energy of the
deuteron ground state (identical for all three substates with 
$\nu=J_{z}=-1,0,1$).
The two-body term $n_{2}$ in this approximation is limited to the
two-nucleon scattering contributions. Hence, no
nucleon-deuteron or deuteron-deuteron correlations are included. 
In the following, we consider only s-wave contributions for simplicity and have
\begin{equation}
\label{eq:n2}
 n_{2}(T,\mu_{p},\mu_{n}) = 
  \frac{2}{\Lambda_{p}^{3}} 
 \exp \left(\frac{2\mu_{p}}{T}\right) b_{pp}
 +  \frac{2}{\Lambda_{n}^{3}} 
 \exp \left(\frac{2\mu_{n}}{T}\right) b_{nn}
 + \frac{2}{\Lambda_{p}^{3/2}\Lambda_{n}^{3/2}} 
 \exp \left(\frac{\mu_{p}+\mu_{n}}{T}\right) b_{pn}
\end{equation}
with the (continuum) virial coefficients (assuming for strong
interactions the symmetry $b_{pp}=b_{nn}$ and $b_{np}=b_{pn}$)
\begin{eqnarray}
\label{eq:bnn}
  b_{nn} & = & 2^{3/2} \int_{0}^{\infty} \frac{dE}{\pi} 
  \exp \left( -\frac{E}{T}\right)\frac{d\delta_{{}^{1}S_{0}}^{(nn)}}{dE} \: ,
 \\
  b_{pn} & = & 2^{1/2} \int_{0}^{\infty} \frac{dE}{\pi} 
  \exp \left( -\frac{E}{T}\right) \left[
    \frac{d\delta_{{}^{1}S_{0}}^{(pn)}}{dE} + 3  \frac{d\delta_{{}^{3}S_{1}}^{(pn)}}{dE} 
 \right] 
\end{eqnarray}
with phase shifts in the isospin triplet ($T=1$, ${}^{1}S_{0}$) and
singlet ($T=0$, ${}^{3}S_{1}$) channels.
The phase shifts can be taken from experiments such that no further
model parameters are needed \cite{HS}. 
Instead one can also use for their calculation a nucleon-nucleon interaction 
potential which is adjusted to describe the empirical scattering phase shifts. 

\subsection{Ambiguity of bound state contributions and physical picture}
\label{Sec:bound}

A partial integration gives the
alternative expressions
\begin{eqnarray}
  b_{nn} & = & \frac{2^{3/2}}{T} \int_{0}^{\infty} \frac{dE}{\pi} 
  \exp \left( -\frac{E}{T}\right) \delta_{{}^{1}S_{0}}^{(nn)}(E) \: ,
 \\
  b_{pn} & = & \frac{2^{1/2}}{T} \int_{0}^{\infty} \frac{dE}{\pi} 
  \exp \left( -\frac{E}{T}\right) \left[
    \delta_{{}^{1}S_{0}}^{(pn)}(E) + 3 \delta_{{}^{3}S_{1}}^{(pn)}(E)
 \right] - 3 \sqrt{2} \: .
\end{eqnarray}
for the virial coefficients, since there is the deuteron bound state
in the ${}^{3}S_{1}$ channel and the Levinson theorem requires
$\delta_{c}(0) = n_{c}\pi$ with the number of bound states $n_{c}$ in channel
$c$. The term $ - 3 \sqrt{2}$ from the lower boundary of the integral 
in $b_{pn} $ can be combined with the bound state contribution $n_{1}^{(d)}$ 
in (\ref{eq:n1}) to give the modified single deuteron contribution 
($\Lambda_{d} \approx \sqrt{\Lambda_{p}\Lambda_{n}/2}$)
\begin{equation}
\label{nd1tilde}
 \tilde{n}_{1}^{(d)} = 
 \frac{3}{\Lambda_{d}^{3}} \exp
 \left(\frac{\mu_{p}+\mu_{n}}{T}\right) 
 \left[  \exp \left(\frac{B_{2,1}}{T}\right) - 1 \right]
\: .
\end{equation}
Both expressions $n_{1}^{(d)}$, Eq. (\ref{nd1}) and $\tilde{n}_{1}^{(d)}$ 
(\ref{nd1tilde}) are related by a partial integration that changes the 
contribution of scattering states in the deuteron channel of Eq.~(\ref{eq:n2}).
Thus we conclude that the contribution of the  scattering states can be reduced
if the bound state contribution is redefined adequately. 
Part of the continuum contribution can be transferred to the bound state 
contribution. 
Therefore the subdivision of the correlated part of the density into a bound 
state and a scattering part is ambiguous.
This problem has been extensively discussed for Coulomb systems where for the 
bound state part the Brillouin-Planck-Larkin partition function has been
introduced to avoid artificial singularities \cite{KKER}. 
As a consequence it makes only sense to consider correlations in a particular 
channel defined by the corresponding quantum numbers without dividing into 
bound and scattering parts. 
Thus we can write
\begin{equation}
 n = n_{\rm free} + n_{\rm corr}
\end{equation}
with the contribution of the free nucleons
\begin{equation}
 n_{\rm free} = n_{1}^{(p)} + n_{1}^{(n)}
\end{equation}
and the contribution of the correlated nucleons
\begin{equation}
 n_{\rm corr} = n_{1}^{(d)} + n_{2} 
 = n_{\rm corr}^{(pp)} + n_{\rm corr}^{(nn)}
 + n_{\rm corr}^{(pn)T=0} + n_{\rm corr}^{(pn)T=1}
\end{equation}
that can be split further into the four nucleon-nucleon channels considered 
above.
This point of view corresponds to the {\em physical picture} where
all many-body states are considered as correlations of nucleons.

\subsection{Continuum correlations and quasiparticle
  shifts}\label{App:Quasip}

Coming back to the generalized Beth-Uhlenbeck formula where the 
single-particle contribution is given by the quasiparticles containing 
the self-energy shifts, the contribution of the continuum states 
(\ref{eq:n2qu}) differs from the simple Beth-Uhlenbeck formula 
(\ref{eq:n2_vir}). The difference (expressed by the $2  \sin^2(\delta_c)$
term) arises because part of the interaction is already accounted 
for in the quasiparticle shifts.
As shown below 
in a model calculation, the Born approximation is fully accounted 
for by the Hartree-Fock (HF) self-energy shifts. Therefore, the 
introduction of the quasiparticle picture reduces the contribution 
of the continuum to the virial coefficient. On the other hand, 
the generalized Beth-Uhlenbeck formula describes the high 
density region as well where the bound state contributions 
are suppressed by Pauli blocking.

The lowest order contribution (Born approximation) of the 
two-body scattering continuum 
to the EoS can be shifted to the contribution of free quasiparticles 
with Hartree-Fock mean-field energy shifts. In the
physical picture with nucleons as the only degrees of freedom, we start from
a non-relativistic Hamiltonian
\begin{equation}
 H = \sum_{1} E(1) a_{1}^{\dagger} a_{1} + \frac{1}{2}
 \sum_{12,1^{\prime}2^{\prime}} V(12,1^{\prime}2^{\prime}) 
 a_{1}^{\dagger}a_{2}^{\dagger} a_{2^{\prime}} a_{1^{\prime}}
\end{equation}
where $\{1\}$ denotes the momentum $\vec{p}_{1}$, spin $\sigma_{1}$
and isospin $\tau_{1}$ characterizing the neutron or proton state. The
kinetic energy is $E(1) = p_{1}^{2}/(2m_{1})$ and the potential energy
contains the matrix element $V(12,1^{\prime}2^{\prime})$ of the
nucleon-nucleon interaction. Sophisticated nucleon-nucleon potentials
can be used that are fitted to available experimental data. 
For demonstration,  
we perform exploratory calculations based on the Yamaguchi separable 
interaction potential \cite{Yamaguchi}, 
\begin{equation}
V(12,1^{\prime}2^{\prime}) = 
 - \frac{\lambda}{\Omega} 
 \frac{\gamma^{2}}{p^{2}+\gamma^{2}} 
 \frac{\gamma^{2}}{(p^{\prime})^2+\gamma^2} \delta_{P,P^{\prime}} 
\delta_{\sigma_{2},\sigma_{2^{\prime}}}\delta_{\tau_{1},\tau_{1^{\prime}}}
 \delta_{\tau_{2},\tau_{2^{\prime}}} \: ,
\end{equation}
with the relative momentum $\vec{p}=(\vec{p}_{2}-\vec{p}_{1})/2$ and the
center-of-mass momentum $\vec{P}= \vec{p}_{1}+\vec{p}_{2}$ (supposing 
$m_{p}=m_{n}$). 
Values for the parameters $\lambda, \gamma$ fitted to nuclear data are given in
Ref.~\cite{Yamaguchi}.

According to the Beth-Uhlenbeck formula, 
the contribution of the $nn$ scattering continuum to the density reads
(c.f. Eqs.~(\ref{eq:n2}) and (\ref{eq:bnn}))
\begin{equation}
 n_{\rm corr}^{(nn)} = 
 \frac{2^{5/2}}{\Lambda_{n}^{3}} \exp \left( \frac{2\mu_{n}}{T}\right)
 \int_{0}^{\infty} \frac{dE}{\pi T} \exp\left(- \frac{E}{T} \right)
 \delta_{{}^{1}S_{0}}(E)~,
\end{equation}
where $E=p^{2}/m_{n}$ is the energy of relative motion.
For the Yamaguchi potential we have in the weak scattering limit 
(Born approximation) the phase shift
\begin{equation}
\delta_{{}^{1}S_{0}}^{(nn)}(E)=  \frac{m_{n} \lambda \gamma}{4 \pi} 
 \frac{\sqrt{\frac{m_{n}E}{\gamma^{2}}}}{\left(1+
 \frac{m_{n}E}{\gamma^{2}}\right)^{2}} \: .
\end{equation}
With this expression we find
\begin{equation}
\label{eq:contcorr}
 n_{\rm corr}^{(nn)} = 
  \frac{2^{3/2}}{\Lambda_{n}^{3}} \exp \left( \frac{2\mu_{n}}{T}\right)
 \frac{\lambda \gamma^{3}}{\pi^{2} T} 
 \int_{0}^{\infty} dx \: \frac{x^{2}}{(1+x^{2})^{2}} 
 \exp\left(- \frac{\gamma^{2}x^{2}}{m_{n}T} \right)
\end{equation}
by introducing $x = \sqrt{m_{n}E/\gamma^{2}}$.

On the other hand, we have from the free quasiparticle 
contribution in lowest order with respect to $\lambda$ 
(Hartree-Fock approximation)
\begin{equation}
 n_{n}^{\rm qu} = 2 \int \frac{d^{3}p_{1}}{(2\pi)^{3}}
 \left[ \exp \left( \frac{p_{1}^{2}/(2m_{n})+\Delta^{\rm
         HF}(p_{1})-\mu_{n}}{T}\right)+1\right]^{-1}
\end{equation}
with the self-energy shift
\begin{equation}
 \Delta^{\rm HF}(p_{1}) = 
 \sum_{p_{2}}
 \left. V(\vec{p}_{1}\vec{p}_{2},\vec{p}_{1}\vec{p}_{2})\right|_{\rm   ex}
 \left[ \exp \left( \frac{p_{2}^{2}/(2m_{n})+\Delta^{\rm HF}(p_{2})
 -\mu_{n}}{T}\right)+1\right]^{-1} ~.
\end{equation}
Without the exchange term, that can be neglected at low densities, and
after expanding for small $\Delta^{\rm HF}$, 
we have in the nondegenerate case 
\begin{equation}
  n_{n}^{\rm qu} =  2 \int \frac{d^{3}p_{1}}{(2\pi)^{3}}
 \exp \left( - \frac{p_{1}^{2}/(2m_{n})-\mu_{n}}{T}\right)
 \left[ 1 - \frac{\Delta^{\rm HF}(p_{1})}{T} + \dots\right]
 = n_{1}^{(n)} + n_{\rm corr}^{(nn)}
\end{equation}
with a term $n_{\rm corr}^{(nn)}$ in addition 
to the free neutron term $n_{1}^{(n)}$. It is given by
\begin{equation}
  n_{\rm corr}^{(nn)}  \approx   2 \int \frac{d^{3}p_{1}}{(2\pi)^{3}}
 \exp \left( - \frac{p_{1}^{2}/(2m_{n})-\mu_{n}}{T}\right)
 \frac{\Omega}{T} \int \frac{d^{3}p_{2}}{(2\pi)^{3}} 
  \frac{\lambda}{\Omega} \left[
 \frac{\gamma^{2}}{(\vec{p}_{2}-\vec{p}_{1})^{2}/4+\gamma^{2}}\right]^{2}
 \exp \left( -\frac{p_{2}^{2}/(2m_{n}) -\mu_{n}}{T}\right) ~.
\end{equation}
Introducing relative and center-of-mass momenta we have with $x=p/\gamma$
\begin{eqnarray}
\label{eq:qucorr}
  n_{\rm corr}^{(nn)}  & \approx &  2 
 \exp \left(  \frac{2\mu_{n}}{T}\right)
 \int \frac{d^{3}P}{(2\pi)^{3}}
 \exp \left( - \frac{P^{2}}{4m_{n}T}\right)
 \frac{\lambda}{T} \int \frac{d^{3}p}{(2\pi)^{3}} 
  \left[
 \frac{\gamma^{2}}{p^{2}+\gamma^{2}}\right]^{2}
 \exp \left( -\frac{p^{2}}{m_{n}T}\right)
 \\ \nonumber & = &
  \frac{2^{3/2}}{\Lambda_{n}^{3}} 
 \exp \left(  \frac{2\mu_{n}}{T}\right)
 \frac{\lambda \gamma^{3}}{\pi^{2}T} \int_{0}^{\infty} dx \: 
 \frac{x^{2}}{(1+x^{2})^{2}}
 \exp \left( -\frac{\gamma^{2}x^{2}}{m_{n}T}\right) \: .
\end{eqnarray}
The expressions (\ref{eq:contcorr})
and  (\ref{eq:qucorr}) coincide in the
lowest order of the interaction strength $\lambda$. 
The factor $2 \left[\sin(\delta_{{}^{1}S_{0}})\right]^{2}$ occurring
in the continuum 
contribution to the density in the generalized Beth-Uhlenbeck formula
(\ref{eq:n2qu}) 
produces a lowest order ${\cal O}(\lambda^3)$ in the weak interaction
limit.  Thus, the lower-order contributions in the continuum part of the
ordinary Beth-Uhlenbeck 
formula are transferred to the single quasiparticle term
$n_{n}^{\rm qu}$ of the density.

\begin{figure}[ht] 
	\includegraphics[width=0.45\textwidth]{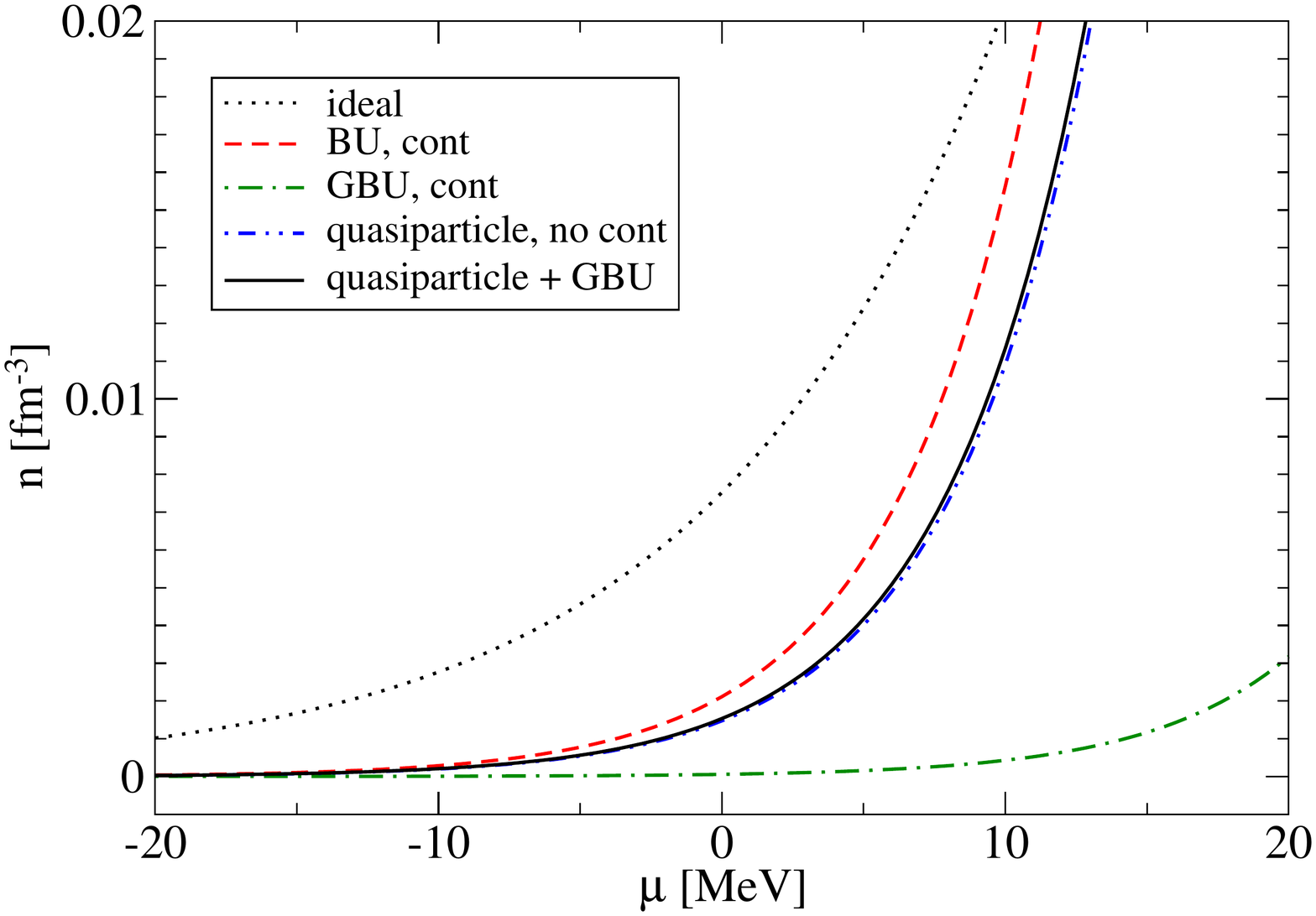}
	\includegraphics[width=0.45\textwidth]{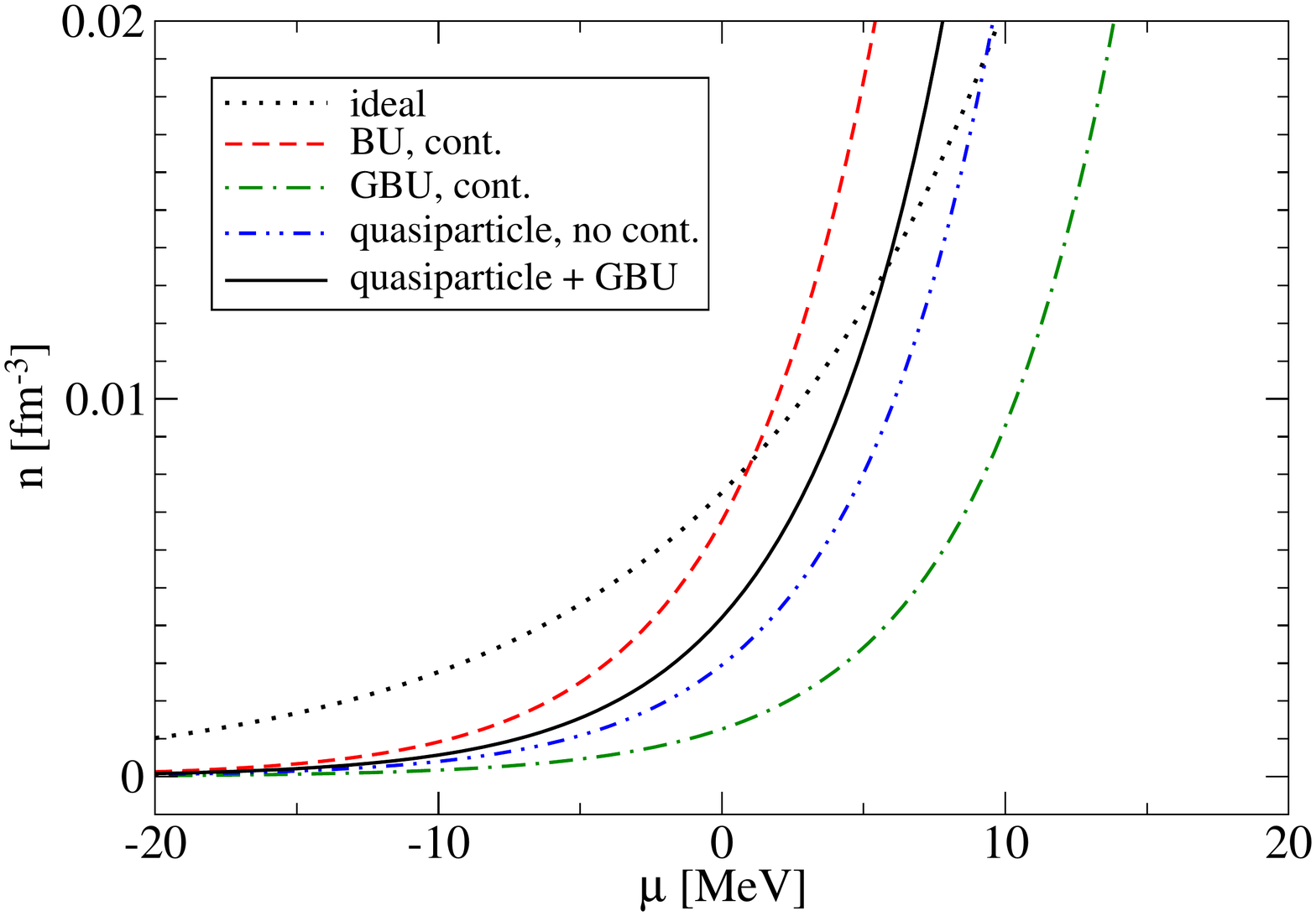}
	\caption{(Color online) Baryon number density $n$ as a function of the 
chemical potential $\mu$ for $T=10$ MeV in neutron matter.
Model calculations for a Yamaguchi interaction with an effective range 
parameter $\gamma = 1.3943\;\mathrm{fm^{-1}}$ are shown for two values of the 
coupling strength: $\lambda=\lambda_0/2$ (left panel) and 
$\lambda=\lambda_0$ (right panel), where $\lambda_0= 704.76$ MeV.
The ideal gas of nucleons (black dotted line) is compared with the correlated 
part obtained from the standard Beth-Uhlenbeck formula (red dashed line),
the generalized Beth-Uhlenbeck formula (green dash-dotted line) containing the 
factor $2\left[\sin\delta\right]^{2}$, the difference in the density if 
quasiparticles are introduced (blue dash-dotted line), and the total 
correction part of density (black solid line).}
 \label{Fig:6} 
 \end{figure}

As an example, calculations have been performed for a
Yamaguchi interaction fitted to $nn$ scattering interaction 
($\gamma = 1.3943$~fm$^{-1}$, $\lambda = 704.76$~MeV~fm$^3$) 
and for a weaker interaction with half the potential strength,
i.e.\ $\lambda /2$, see Fig.~\ref{Fig:6} with the total baryon
number density as a function of the chemical potential $\mu=\mu_{n} $ 
(neutron matter) in different approximations.
The correlated part of the density, 
that is contained in the quasiparticle picture, almost completely reproduces 
the second virial coefficient in the case of lower interaction strength. 
For stronger interaction the quasiparticle shift accounts only for a part of 
the second virial coefficient.
Note that the $nn$ interaction is strong, and the di-neutron is almost bound. 
Therefore, a mean-field approach is not sufficient to account for the continuum
contributions.

Neutron matter contains no clusters because the interaction is not strong 
enough to form a bound state. 
Instead of the cluster virial expansion, we have the standard virial expansion 
where the second virial coefficient as a benchmark can be directly related to 
the observed phase shifts. 
For details see \cite{HSn,VT}. The introduction of quasiparticle
states and the reduction of the scattering contributions according 
to the generalized Beth-Uhlenbeck approach \cite{SRS} is an important issue to 
go to higher densities. 

\section{Quantum statistical approach and  the cluster virial expansion}
\label{Sec:4}

A systematic approach to the cluster expansion of thermodynamic
properties is obtained from quantum statistics. 
The grand canonical thermodynamic potential 
\begin{equation}
 J=-P \Omega=-T \ln {\rm Tr}\,\, {\rm e}^{-(H-\mu N)/T},
\end{equation}
where $P$ is the pressure and $\Omega$ the volume, can be represented 
by diagrams within a perturbation expansion \cite{KB}, see also \cite{KKER}. 
We have
\begin{equation}
 P= \frac{1}{\Omega} {\rm Tr} \ln[-G_1^{(0)}]-\frac{1}{2 \Omega} \int_0^1 
\frac{d \lambda}{\lambda}{\rm Tr} \Sigma_\lambda G_\lambda,
\end{equation}
or
\centerline{\includegraphics[width=0.6\textwidth]{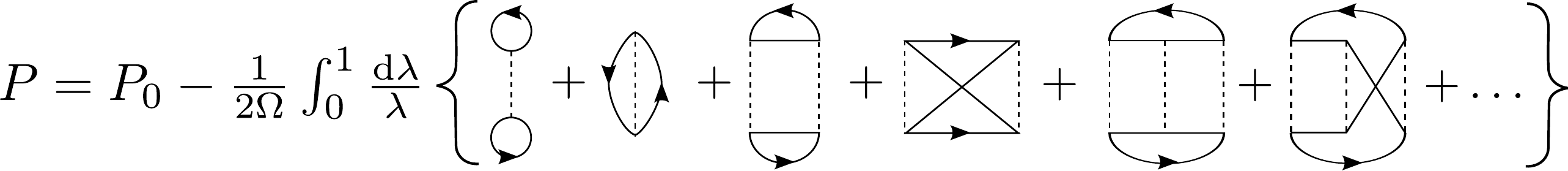}}
\vspace{0.5cm}
\noindent
where $\lambda$ is a scaling factor substituting the interaction $V$ by 
$\lambda V$.
$G_1^{(0)}$ is the free single-particle propagator that gives the ideal part 
of the pressure $P_0$.
The full single-particle Green function $G_\lambda$ and the self-energy 
$\Sigma_\lambda$ are taken with the coupling constant $\lambda$.
Depending on the selected diagrams, different approximations can be found. 
In particular, the second virial coefficient for charged particle systems has 
been investigated, see Ref.~\cite{KKER}.
\footnote{Note that the interaction in nuclear systems is strong. 
However, the perturbation expansion is performed with respect to the imaginary 
part of the self-energy that is assumed to be small. 
Most of the interaction is already taken into account in the self-consistent 
determination of the quasiparticle energies. 
With increasing density, the Fermi energy will dominate the potential energy 
so that the correlations are suppressed.
A quasiparticle description can be used to calculate the nuclear structure.
}

An alternative way to derive the equation of state is to start from the 
expression for the total nucleon  density
\begin{equation}
n_{\tau_1}(T,\mu_p,\mu_n)=\frac{2 }{ \Omega} \sum_{1}
\int_{-\infty}^\infty\frac{d \omega}{2 \pi} f_{1}(\omega) S_1(1,\omega)\,, 
\label{eosspec}
\end{equation}
where $\Omega$ is the system volume, $\tau_1 =n,\,p$, and summation over spin 
direction is collected in the factor 2. 
Both the Fermi distribution function and the spectral function depend on the 
temperature and the chemical potentials $\mu_p, \mu_n$ not given explicitly.
The spectral function $S_1(1,\omega)$ of the single-particle Green function 
$G_1(1,i z_\nu)$ is related to the single-particle self-energy $\Sigma(1,z)$ 
according to
\begin{equation}
\label{spectral}
 S_1(1,\omega) = {2 {\rm Im}\,\Sigma_1(1,\omega-i0) 
\over(\omega - E(1)- {\rm Re}\, \Sigma_1(1,\omega))^2 + 
({\rm Im}\,\Sigma_1(1,\omega-i0))^2 }\,,
\end{equation}
where the imaginary part has to be taken for a small negative
imaginary part in the frequency.

Both approaches are equivalent. As shown by Baym and Kadanoff \cite{KB}, 
self-consistent approximations to the one-particle Green function can be given 
based on a functional $\Phi$ so that
\begin{equation}
\Sigma_1(1,1') = \frac{\delta  \Phi}{\delta G_1(1,1')}.
\end{equation}
Different approximations for the generating functional $\Phi$ are discussed in 
the App. \ref{App:4}.
The self-consistent $\Phi$-derivable approximations not only lead to a 
fully-conserving transport theory.
In the equilibrium case they also have the property that different methods to 
obtain the grand partition function such as integrating the expectation value 
of the potential energy with respect to the coupling constant $\lambda$, or 
integrating the density $n$ with respect to the chemical potential $\mu$, lead 
to the same result. 
In particular, with
\begin{equation}
 J = -{\rm Tr}\,\, \ln (-G_1) -{\rm Tr} \Sigma_1 G_1 +\Phi
\end{equation}
also 
\begin{equation}
 n = - \frac{1}{\Omega} \frac{\partial J}{\partial \mu}
\end{equation}
holds in the considered approximation.

\begin{figure}[ht] 
 \centerline{\includegraphics[width=0.5\textwidth]{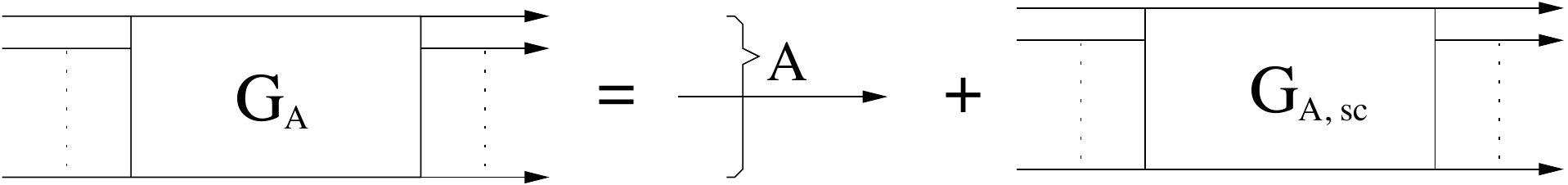}} 
 \caption{Splitting of the $A$-particle cluster propagator into a 
   bound and scattering contribution. Note that the internal quantum number has
been dropped.
   } 
 \label{Fig:2} 
 \end{figure}

The latter approach using Eq. (\ref{eosspec}) has been extensively used in 
many-particle systems \cite{Zimmermann:1985ji,RMS,SRS}, in particular in 
connection with the chemical picture. 
An analysis of the self-energy allows to work out a diagram technique that 
treats bound states on the same footing as ``elementary'' single particle 
described by the free propagator
\begin{equation}
\label{G1}
 G_1^{(0)}(1,z) = \frac{1}{z-E_1(p_1)}\,.
\end{equation}
Considering the $A$-particle propagator (\ref{bilinear}) in the low-density 
limit where we can drop all medium effects, the solution of the Bethe-Salpeter 
equation in ladder approximation gives the propagator for the $A$-particle 
bound states
\begin{eqnarray}
 G_{A,\nu}^{\rm bound}(1\dots A;1'\dots A';z_A)&=&
 \langle 1\dots A |\psi_{A \nu P} \rangle \frac{1}{z_A-E^{(0)}_{A,\nu}(P)} 
 \langle \psi_{A \nu P} |1'\dots A' \rangle\,
\label{Apropagator}
\end{eqnarray}
where $\nu$ indicates the internal quantum state of the $A$-particle bound 
state, after separation of the center-of-mass momentum $\vec P$.
As a new element, the bound state propagator is introduced as indicated in 
Fig.~\ref{Fig:2}. 
This bound state propagator has the same analytical form like the single 
particle propagator (\ref{G1}), besides the appearence of the internal wave 
function that determines the vertex function.

As example, in App. \ref{App:4} different approximations are obtained such as 
the nuclear statistical equilibrium (NSE) and the cluster mean-field (CMF) 
approximation using the chemical picture.
These approximations are based on the bound state part of the $A$-particle 
propagator.
They give leading contributions in the low-density, low temperature range where
bound states dominate the composition of the many-particle system.
From the point of view of the physical picture, these contributions arise in 
higher orders of the virial expansion of the equation of state. 
As example, the formation of the $A$-particle bound state 
is seen in the $A$-th virial coefficient, the mean-field shift due to a 
cluster $B$ in the $(A+B)$-th
virial coefficient. The chemical picture indicates which high-order virial 
coefficients of the virial expansion are essential, if the many-particle 
system is strongly correlated so that bound states are formed.

As an example, Hydrogen molecules dominate the electron-proton system under 
normal conditions.
The effective interaction can be calculated, and the corresponding virial 
coefficient determines the non-ideal part of the Hydrogen EoS. 
Local effective interactions are the Lennard-Jones or the Morse potential, 
and the Mayer cluster expansion can be used to evaluate the Hydrogen-Hydrogen 
virial coefficient.
Within the physical picture based on electrons and protons, the eighth-order 
virial coefficient has to be analyzed to find the non-ideal features of the 
Hydrogen gas.
Similar approaches can also be used for $\alpha$ matter in nuclear physics, 
introducing effective interactions such as the Ali-Bodmer potential.

Using the quasiparticle approach, the  EoS, Eq. (\ref{genclBU}),
is obtained, see App.~\ref{App:cese}. In contrast to the
NSE, Eq. (\ref{eq:idgas}),
medium dependent quasiparticle energies are used. 
In the cluster mean-field (CMF) approximation only the first-order terms of 
the cluster-cluster interaction are taken into account.
The remaining part of the continuum correlations is neglected. 

To improve the approximation, the scattering part of the $A$-particle 
propagator has to be considered.
It contributes also to the $A$-th virial coefficient. 
The scattering processes within the $A$-particle system can have different 
channels.
As an example we discuss here binary elastic scattering processes between 
sub-clusters $A_1$ and $A_2$ of the system of $A$ particles, $A=A_1+A_2$. 
Binary phase shifts $\delta_{A_1,A_2}(E)$ are introduced that describe the 
corresponding scattering experiments. 
They can also be calculated within few-body theory. 
Besides the effective interaction between the sub-clusters that are depending 
on the internal wave function of the sub-clusters, also virtual transitions to 
excited states have to be taken into account.
In general, the effective interaction is non-local in space and time, i.e.\ 
momentum and frequency dependent.

The generalized  cluster Beth-Uhlenbeck formula (\ref{eq:n2qu}) is obtained 
when in particle loops not the free propagator, but 
quasiparticle Green's functions are used.
If the quasiparticle shift is calculated in Hartree-Fock
approximation,
the first order term of the interaction must be excluded from the ladder
$T_2^{\rm ladder}$ 
matrix to avoid double counting. The bound state part is not affected, 
it is determined by an infinite number of diagrams. The scattering part 
is reduced subtracting the Born contribution as shown in
 Eq.\ (\ref{eq:n2qu}) 
by the $2 \left[\sin (\delta_{c})\right]^{2}$ term; for the derivation
see Ref. \cite{SRS}.

The continuum correlations that are not considered in the NSE give a 
contribution to the second virial coefficient in the chemical picture. 
We can extract from the continuum part two contributions: 
resonances that can be treated like new particles in the law of mass action, 
and the quasiparticle shift of the different components contributing
to the law of mass action. 
Both processes are expected to represent significant contributions of the 
continuum. 
After projecting out these effects, the residual contribution 
of the two-nucleon continuum is assumed to be reduced. 
One can try to parametrize the residual part, using the ambiguity in 
defining the bound state contribution, see Sec. \ref{Sec:bound}. 
Eventually the residual part of the continuum correlations can be neglected.

\section{Conclusions}
\label{Sec:5}

We have shown that it is possible to give an unified description for
the nuclear matter equation of state that joins the region of
saturation density, where quasiparticle approaches can be used, 
with the low-density region where the  Nuclear Statistical
Equilibrium 
or, more rigorously, the virial expansion can be applied. 
The chemical picture that allows for a systematic treatment of 
all bound states is used as guide line to formulate cluster expansions 
of different quantities such as self-energy, density and pressure. 
The inclusion of arbitrary nuclei as demanded by the chemical picture 
is indispensable to derive the nuclear matter EoS.

As a main ingredient, the generalized Beth-Uhlenbeck formula
\cite{RMS,SRS} that treats the two-particle correlations
already joins  the low-density limit with the saturation density
region. It is extended to clusters with arbitrary mass number  
\cite{RMS} and has been investigated recently \cite{unsere} 
to derive the thermodynamic properties of nuclear matter in the subsaturation 
density region.

Whereas the bound states are treated in a systematic way, 
the continuum of scattering states needs further investigations, 
especially if many channels appear in the continuum. 
The approach presented is to extract important contributions 
from the continuum that are of relevance for the physical properties. 
We discussed three contributions:\\
 i) Resonances in the continuum 
can be considered similar to bound states. As example, 
we can consider $A=8$ and the subdivision into two $\alpha$ particle
clusters.  The $\alpha-\alpha$ elastic scattering phase shifts 
can be used to find a contribution to the corresponding 
cluster-virial coefficient. $^8$Be as a sharp resonance can be 
treated as new particle in the chemical picture and should be 
projected out from the scattering contribution to the equation of state. \\
ii) Mean-field effects can be extracted introducing quasiparticles 
not only for the single nucleon states, but also for the bound
states.  This reduces the interaction contribution to the 
continuum states. We have shown that in the weak scattering limit 
the continuum contributions are transferred to the quasiparticle 
energy shifts. In particular, the generalized Beth-Uhlenbeck formula 
shows this behavior. \\
iii) There is no first principle distinction 
between the contribution of bound states and scattering states 
to the virial coefficient. For instance, partial integration and 
using the Levinson theorem gives different expressions for the
contribution of bound states, see Sec.\ \ref{Sec:bound}. 
The contribution of the bound states to the virial coefficient 
is not unambiguously defined. This allows to find optimal
approximations to take the contribution of the continuum into account.

In conclusion, the cluster-virial expansion in combination with 
the consideration of excited states, resonances, and the introduction
of the quasiparticle concept allows to extract relevant contributions 
from the continuum states. The remaining contributions of the 
continuum of the cluster-virial coefficients can be included 
into the sum over the internal quantum number $\nu$ in the respective cluster-cluster channel. 
A systematic approach to the residual contribution that avoids 
double-counting is possible starting from the generating functional, 
but needs further  study including investigations of
topological aspects to characterize sets of relevant diagrams 
within the perturbation expansion. 
This is a complex problem and has to be worked out in future. 
Alternatively one can also use numerical methods to simulate
the many-nucleon systems in the intermediate region. 
This way the interpolation between the saturation density and the low-density 
limit can be improved.

Improved calculations of the nuclear matter EoS take the contribution of the 
continuum to the second virial coefficient into account. 
The ambiguity in the definition of the bound state contribution has been used
in quantum statistical calculations, see the quantum statistical calculation 
in Ref.\ \cite{unsere} and the recent work \cite{VT}. Formally, the summation 
over the internal quantum number $\nu$ can be used to consider also the
contribution of scattering states, as well-known from the Planck-Larkin 
partition function in plasma physics \cite{KKER}.
The effect of the correct treatment of continuum correlations for calculating 
the composition of nuclear matter is shown in Ref.\ \cite{unsere}. 
The dissolution of clusters with increasing density was calculated within
two different approaches, the quantum statistical approach where the 
second virial coefficient was taken into account, and the generalized RMF
approach where continuum contributions are neglected. The largest 
discrepancies are obtained for the deuteron fraction at high temperatures
when the deuteron binding energy is small compared with the temperature
(see also \cite{VT}). The influence of the correct treatment of continuum 
correlations in the deuteron channel on other thermodynamic quantities
can also be seen when comparing both approaches. 
The systematic inclusion of further scattering phase shifts, e.g., in the 
$\alpha-\alpha$ channel, within the cluster virial expansion would give an 
improved approach to all thermodynamic quantities.

\begin{acknowledgments}
This work has been supported by CompStar, a Research Networking Programme 
of the European Science Foundation and by the Polish Ministry for Science and 
Higher Education supporting this network. 
S.T. was supported by the Helmholtz Association (HGF) through the 
Nuclear Astrophysics Virtual Institute (VH-VI-417). 
D.B. has been supported by the Polish Narodowe Centrum Nauki (NCN) under grant 
No. NN 202 231837 and by the Russian Fund for Basic Research (RFBR) under grant
No. 11-02-01538-a. 
The work of T.K. is supported by the ``hadronphysics3'' network within the 
seventh framework program of the European Union.  
We acknowledge the support by the Extreme Matter Institute (EMMI), 
and by the DFG cluster of excellence ``Origin and Structure of the Universe''.
\end{acknowledgments}

\appendix

\section{Cluster expansion of the single-nucleon self-energy}
\label{App:cese}

To include nuclei with arbitrary mass number $A$, a cluster
decomposition of the single-nucleon self-energy can be performed \cite{RMS}.
In particular,  the inclusion of the light elements $^3$H, $^3$He, 
and $^4$He has been discussed in \cite{unsere}.
We will not repeat the Green function approach here, see, 
e.g., Refs. \cite{RMS,R2011,unsere} but give only some final results. 
The nucleon density is expressed in terms of the spectral function. 
The latter is related to the single-nucleon self-energy that 
is represented by Feynman diagrams.
According to the chemical picture, the single nucleon propagators 
that occur in the self-energy have to replaced by propagators of 
arbitrary clusters. The cluster decomposition of the self-energy 
yields the law of mass action and the NSE.
The mean-field approximation has to be replaced by a cluster 
mean-field approximation as given in App.~\ref{App:CMF}. 
This way the quasiparticle concept for the single-nucleon state is extended to 
arbitrary clusters.

The evaluation of the self-energy gives
\begin{equation}
\label{SigmaT}
 \Sigma (1, z_\nu) = \sum_A \sum_{z_A, 2...A} G^{(0)}_{(A-1)}(2,...,A,i z_A - iz_\nu) T_A(1...A,1'...A', z_A)
\end{equation}
with the free $(A-1)$ (quasi-) particle propagator 
\begin{equation}
\label{G0quasi}
G^{(0)}_{(A-1)}(2,...,A, z) = \frac{1}{z-E_2-...-E_A} \frac{f_{1,Z_2}(2)...f_{1,Z_A}(A)}{f_{A-1,Z_{A-1}}(E_2+...+E_A)}\,.
\end{equation}
The quantity $z_\nu$ is the single-particle Matsubara frequency, and $z_A$ that of the $A$-particle system.
The $A$-particle T matrix is obtained from the $A$-particle Green function by amputation.

The $A$-particle Green function obeys in ladder approximation a Bethe-Salpeter equation (BSE)
\begin{eqnarray}
&&G_A(1...A,1'\dots A',z_A)=G^{(0)}_A(1...A,z_A)
\delta_{11'}\dots\delta_{AA'} \nonumber \\ && + \sum_{1''\dots A''}G^{(0)}_A(1...A,z_A) 
V^{A,{\rm mf}}(1...A,1''\dots A'')G_A(1''...A'',1'\dots A',z_A)
\label{BSE}
\end{eqnarray}
where $ V^{A,{\rm mf}}(1...A,1'\dots A')= \sum_{i<j}[V_{ij}+\Delta
V_{ij}^A] $ is the interaction within the $A$-particle cluster,
including mean-field contributions, see App.~\ref{App:CMF}.
The free $A$-quasiparticle Green function results as
\begin{equation}
G^{(0)}_A(1...A,z_A)=\frac{[1-\tilde f_1(1)]\dots [1-\tilde
  f_1(A)]-\tilde f_1(1)\dots \tilde f_1(A)}{z_A - E^{\rm
    qu}_1(1)-\dots- E_1^{\rm qu}(A)} \: .
\label{freeGF}
\end{equation}

The approximation of an
uncorrelated medium, see App.~\ref{App:RMF}, leads to the effective occupation numbers 
\begin{equation}
\label{f1}
\tilde f_1(1) = \frac{1}{\exp[E_1^{\rm
qu}(1)/T- \tilde \mu_\tau/T] +1} \approx \frac{n_\tau}{2} \left(\frac{2 \pi \hbar^2}{m T}\right)^{3/2} 
e^{-E_1^{\rm qu}(1)/T}
\end{equation}
in the low-density, non-degenerate limit ($\tilde \mu_\tau <0  $), where  $\tilde \mu_\tau$ is
determined by the normalization condition $2 \sum_p \tilde f_1(p) =
n_{\tau}$, where $\tau $ denotes isospin (proton or neutron).

The solution of the BSE is given by an expansion, bilinear in the wave 
functions
\begin{equation}
G_A(1...A,1'\dots A',z_A)=\sum_{\nu P}\psi_{A \nu P}(1\dots A)
\frac{1}{z_A-E^{\rm qu}_{A \nu P}} \psi^*_{A \nu P}(1'\dots A')\,.
\label{bilinear}
\end{equation}
The summation over the internal quantum states $\nu$ includes besides 
the bound states also the scattering states. The $A$-particle wave function 
and the corresponding eigenvalues follow from solving the in-medium
Schr\"odinger equation  
\begin{eqnarray}
&&[E_1^{\rm qu}(1)+\dots + E_1^{\rm qu}(A) - E^{\rm qu}_{A \nu}(P)]\psi_{A \nu P}(1\dots,k,\dots A)
\nonumber \\ &&
+\sum_{1'\dots A'}\sum_{i<j}[1-\tilde f_1(i)- \tilde f_1(j)]V(ij,i'j')\prod_{k \neq 
  i,j} \delta_{kk'}\psi_{A \nu P}(1'\dots,k',\dots A')=0\,.
\label{waveA}
\end{eqnarray}
This equation contains the effects of the medium in the quasiparticle shift 
as well as in the Pauli blocking terms. 
Obviously the bound state wave functions and energy eigenvalues as
well as the scattering phase shifts become dependent on the
c.m. momentum $P$, temperature $T$,
and densities $n_p,n_n$. 

Two effects have to be considered, the quasiparticle energy shift and the 
Pauli blocking. 
Detailed results have been obtained for the 
two-nucleon case. The shift of
the binding energy and the medium modification of the scattering phase
shifts are discussed extensively, see \cite{SRS,alm}.
The solutions of the in-medium Schr\"odinger equation (\ref{waveA})
for $A=2,3,4$ was parametrized recently \cite{R2011}.

\section{The Cluster-mean field (CMF) approximation}
\label{App:CMF}

The chemical picture gives the motivation to extend the 
mean-field approximation for the case of cluster formation.
Bound states are considered as new species, to be treated on the same
level as free particles. A conserving mean-field approach can be formulated 
by specifying the Feynman diagrams that are taken into account when treating
$A$-particle cluster propagation \cite{cmf}. The corresponding  
$A$-particle cluster self-energy is treated to first order in 
the interaction with the single particles as well as with 
the $B$-particle cluster states in the medium, but with
full anti-symmetrization between both clusters $A$ and $B$. We use the
notation $\{A,\nu,P\}$ for the particle number, internal quantum number 
(including proton number $Z$)
and center of mass momentum for the cluster under consideration and
$\{B,\bar \nu,\bar P\}$ for a cluster of the surrounding medium.

For the $A$-particle
problem, the effective wave equation reads
\begin{eqnarray}
&&[E(1)+ \dots E(A) - E_{A \nu P}] \psi_{A \nu P}(1 \dots A)\nonumber\\
&& + \sum_{1'\dots A'}
\sum_{i<j}^A V_{ij}^A(1\dots A, 1'\dots A')  \psi_{A \nu P}(1' \dots A')
\nonumber\\ && + \sum_{1'\dots A'}
V_{\rm nm}^{A,{\rm mf}}(1\dots A, 1'\dots A')  \psi_{A \nu P}(1'
\dots A') = 0 \,,
\end{eqnarray}
with $V_{ij}^A(1\dots A, 1'\dots A') = V(12,1'2') \delta_{33'} \dots
\delta_{AA'}$.  The effective potential $V_{\rm nm}^{A,{\rm mf}}
(1\dots A, 1'\dots A')$ describes the influence of the nuclear
medium on the cluster bound states and has the form
\begin{equation}
V_{\rm nm}^{A,{\rm mf}}(1\dots A, 1'\dots A') = \sum_i \Delta
(i) \delta_{11'} \dots \delta_{AA'} + {\sum_{i,j}}' \Delta
V_{ij}^A(1\dots A, 1'\dots A') \,,
\end{equation}
with
\begin{eqnarray}
&&\Delta(1) = \sum_2(V(12,12)_{\rm ex} \tilde f(2) -  \sum^\infty_{B=2}
\sum_{\nu \bar P} \sum_{2 \dots B} \sum_{1' \dots B'} f_B(E_{B  \bar \nu  \bar P})
\times \nonumber\\ && \qquad \qquad \qquad \times 
\sum_{i<j}^m V_{ij}^B(1\dots B, 1'\dots B') \psi_{B  \bar \nu  \bar
  P}(1 \dots B) 
 \psi^*_{B  \bar \nu  \bar P}(1' \dots B')\,, \nonumber
\end{eqnarray}
\begin{eqnarray}
&&\Delta V^A_{12}(1\dots A, 1'\dots A') = - \Biggl\{\frac{1}{2}(\tilde
  f(1) + \tilde f(1')) 
V(12,1'2') +\\
&&\qquad \qquad +\sum_{B=2}^\infty \sum_{\nu \bar P} \sum_{\bar 2 \dots \bar
  B} \sum_{\bar 
  2'\dots \bar B'} f_B(E_{B \bar \nu \bar P})   
\sum_j^B V_{1j}^B (1 \bar2' \dots \bar B', 1' \bar2 \dots
\bar B) 
\times  \nonumber\\ && \qquad \qquad \qquad \times 
\psi^*_{B \bar \nu \bar P} (2 \bar2 \dots \bar B) \psi_{B \bar \nu \bar P} (2' \bar2'
\dots \bar B') \Biggr\} \delta_{33'} \dots \delta_{AA'}\,, \nonumber
\end{eqnarray}
\begin{equation}
\tilde f(1) = f_1(1) + \sum^\infty_{B=2} \sum_{\nu \bar P} \sum_{2 \dots B}
f_B(E_{B  \bar \nu  \bar P}) |\psi_{B \bar \nu \bar P}(1 \dots B)|^2\,,
\end{equation}
where (see Eq.\ (\ref{vert}); the charge number $Z$ counts as internal quantum number)
\begin{equation}
f_A(E)= \frac{1}{e^{  (E- Z \mu_p-(A-Z)\mu_n)/T} - (-1)^{A}} \,\,.
\end{equation}
We note that within the mean-field approximation, the effective potential 
$V_{\rm nm}^{A,{\rm mf}}$ remains energy independent,
i.e.\ instantaneous.  The quantity $\tilde f(1)$ describes the effective
occupation of state $1$ due to free and bound states, while exchange 
is included by the additional terms in $\Delta V^A_{12}$
and $\Delta(1)$, thus accounting for antisymmetrization.

Of course, the self-consistent solution of the cluster in a clustered
medium is a rather involved problem which has not been solved until
now. In particular, the composition of the medium has to be
determined, with energy shifts of the different
components (clusters of $B$ nucleons) in the medium
solving the effective wave equation for the $B$-nucleon problem.

Two effects have to be considered on the single nucleon level, the 
quasiparticle energy shift and the Pauli blocking. 
Phase space is also occupied by clusters as expressed by $\tilde f(1)$. 
This effective occupation number is normalized to the total nucleon density.
As approximation, a Fermi distribution function that is normalized 
correspondingly has been used in recent works \cite{unsere}.
Obviously the bound state wave functions and energy eigenvalues as
well as the scattering phase shifts become dependent on temperature
and density.

\section{Parametrization of single-nucleon quasiparticle shifts}
\label{App:RMF}

Different approaches to determine the single-nucleon quasiparticle self-energy 
shifts are known from the literature. 
We will not give an exhaustive review but mention only some general features, 
see also \cite{Klahn:2006ir}. 
There are first principle approaches that start from  realistic nucleon-nucleon
interactions and solve the many-particle problem numerically. 
Quasiparticle self-energy shifts  can also be obtained by different approaches 
such as the Dirac-Brueckner Hartree-Fock methods or phenomenological density 
functionals. 
The latter are based on effective density dependent interactions 
such as Gogny or Skyrme forces or use relativistic mean-field (RMF) concepts, 
see \cite{Typel1999,unsere,Gaitanos,Baran}.
Parameter are adjusted to reproduce nuclear bulk properties and properties of 
nuclei.

There are several relativistic mean-field parametrizations actually used, 
like TW99 \cite{Typel1999},
DD \cite{Typel2005}, 
DD2 \cite{unsere}, 
TM1 \cite{Shen1998,Shen2011hyperon}, 
TMA \cite{Hempel2010}, 
FSUGold \cite{Todd2005} and 
DDME$\delta$ \cite{RocaMaza2011}.
We focus on the density dependent DD2 parametrization \cite{unsere} 
adjusted to experimental properties of nuclei. 
In particular, this parametrization predicts reasonable values for the 
saturation density ($n_{\rm sat}=0.149$ fm$^{-3}$),  binding energy 
($E/A=-16.02$ MeV), compressibility ($K=242.7$ MeV, what is not far from the 
experimentally determined value of 231 $\pm$ 5 MeV \cite{Youngblood1999}), 
symmetry energy ($J=32.73$ MeV, what fits the experimental value 31.3 MeV 
\cite{Li2010}), and the symmetry slope parameter ($L=57.94$ MeV  consistent
with recent experimental constraints \cite{LimLattimer}).
For neutron stars the model predict a maximum mass of $2.44\,M_\odot$, 
not in conflict with the largest known mass of $M = 1.97\pm0.04\,M_\odot$ 
\cite{Demorest2010}.

For direct use, a parametrization for the DD model \cite{Typel2005} was 
presented in Ref.~\cite{unsere}.
We give here an improved parametrization of the DD2
  model \cite{unsere} in form of a Pad\'e approximation.
The variables are temperature $T$, baryon number
density $n=n_n+n_p$, and the asymmetry parameter  $\delta=1-2Y_p$  with the 
total proton fraction $Y_p=n_p/n$.
The intended relative accuracy in the parameter value range 
$T < 20$ MeV, $n< 0.16$ fm$^{-3}$ is 0.001.

The spectral function in the RMF approach gives the quasiparticle dispersion 
relation ($i=p,n$, no antiparticles) for the (non-relativistic) single 
quasiparticle energies
\begin{equation}
e_i(k) = \sqrt{\left[m_i-S(n, \delta,T)\right]^2+k^2}+V_i(n, \delta,T)
- m_{i} \: .
\end{equation}
In the non-relativistic case we have  
\begin{equation}
 e_i(k) = \frac{k^2}{2[ m_i- S(n,\delta,T)]}+V_i(n, \delta,T) - S(n,\delta,T)~.
\end{equation}
With the quasiparticle energies, the chemical potentials $\mu_{i}$ follow from 
solving
\begin{equation}
n_i = \frac{1}{\pi^2} \int_0^\infty dk \: \frac{k^2}{\exp\left\{ [e_i(k) - 
\mu_i]/T\right\} +1}\,.
\end{equation}

The scalar self-energy (identical for neutrons and
  protons) is approximated as
\begin{equation}
	S(n, \delta,T) = s_0(\delta,T)\; n \frac{1 + s_1(\delta,T) \;n
+  s_2(\delta,T)\;n^2}{1 + s_3(\delta,T)\;n + s_4(\delta,T)\; n^2}
\end{equation}
with coefficients
\begin{equation}
  s_i = s_{i,0} + s_{i,1}\; T +  s_{i,2}\;\delta^2 +  s_{i,3}\;\delta^4\: ,
\end{equation}
densities $n$ in fm$^{-3}$ and temperatures $T$ as well as the self-energies 
$S,V$ in MeV. Parameter values are given in the Table \ref{tab:S}.
\begin{table}[ht]
\begin{tabular}{|c|c|c|c|c|c|}
\hline
$s_{i,j}$& $i=0$	& $i=1$		& $i=2$		& $i=3$		& $i=4$\\
\hline
$j=0$	& $4463.117$   & $20.56456$   & $15.98022$   & $24.27416$   & $114.5972$  \\
$j=1$	& $-6.609841$     & $-0.040985$   & $0.866352$    & $-0.074176$   & $1.349746$    \\
$j=2$	& $-0.170252$     & $-0.339370$   & $-2.020097$   & $-0.542662$   & $2.674353$    \\
$j=3$	& $4.111559$      & $0.997156$    & $-3.018041$   & $1.196491$    & $0.726793$    \\
\hline
\end{tabular}
\caption{\label{tab:S}
Coefficients $s_{i,j}$ for the Pad\'e approximation of the scalar
self-energy $S(n,\delta,T)$.}
\end{table}

The vector self-energy $V_p(n, \delta,T)=V_n(n, -\delta,T)$ is
approximated as
\begin{equation}
	V_p(n, \delta,T) = v_0(\delta,T)\;n  \frac{1 + v_1(\delta,T)\;n
          + v_2(\delta,T)\;n^2}{1 +v_3(\delta,T)\;n + v_4(\delta,T)\;n^2}\\
\end{equation}
with coefficients
\begin{equation}
	v_i = v_{i,0} + v_{i,1}\; T + v_{i,2}\;\delta + v_{i,3}\;\delta^2
\end{equation}
and parameter values given in Table \ref{tab:V}.

\begin{table}[ht]
\begin{tabular}{|c|c|c|c|c|c|}
\hline
$v_{i,j}$& $i=0$	& $i=1$		& $i=2$		& $i=3$		& $i=4$\\
\hline
$j=0$	& $3403.144$   & $0.662946$    & $10.77796$   & $3.432703$    & $23.01450$\\
$j=1$	& $0.000052$      & $-0.006142$   & $0.004432$    & $0.000104$    & $-0.033018$\\
$j=2$	& $486.581687$    & $1.140795$    & $-0.802040$   & $1.548693$    & $5.922645$\\
$j=3$	& $-2.420361$     & $-0.717645$   & $0.457561$    & $-0.336038$   & $0.050892$\\
\hline
\end{tabular}
\caption{\label{tab:V}%
Coefficients $v_{i,j}$ for the Pad\'e approximation of the vector
  self-energy $V_{p}(n,\delta,T)$.}
\end{table}

\section{Generating functional approach for the cluster virial
  expansion}
\label{App:4}
In this Appendix we  present instructive examples of approximations to the 
$\Phi$ functional on the level of Feynman diagrams to demonstrate the power of 
this approach. 
As shown by Baym and Kadanoff [21], any choice of a subset of diagrams for 
$\Phi$ defines a selfconsistent approximation, where every Greens function is 
dressed by its appropriate self-energy obtained by the variation of $\Phi$. 
The advantage of such so-called conserving or $\Phi$-derivable approximations 
is that they satisfy conservation laws and guarantee thermodynamic consistency.
 
In the past the approach has been applied to a wide variety of many-body 
problems \cite{David1,David2,David3,David4,David5,David6} but not yet to the 
problem of cluster formation in nuclear matter which we are going to discuss 
here.

\begin{figure}[ht] 
	\centerline{\includegraphics[width=0.6\textwidth]{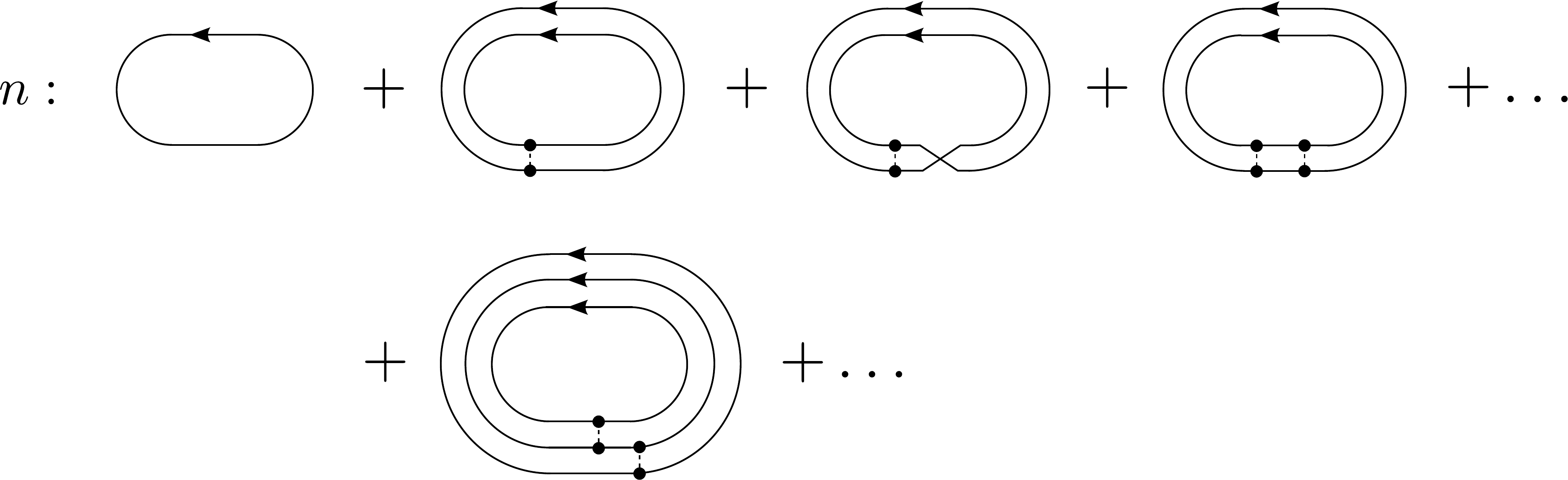}} 
	\caption{Cluster decomposition of the generating functional
          $\Phi$. Some contributions for $A=2,3$ are shown explicitly, 
          the ladder sum has to be
          completed considering an arbitrary number of interaction lines. 
	  The single particle loop has been added to represent the ideal 
	  part of the density.} 
	\label{Fig:1} 
\end{figure}

In the physical picture the generating functional is represented by the sum of 
all diagrams that consist of $A=2, 3, ....$ free fermion loops connected by 
arbitrary numbers of interaction lines. Exchange diagrams have to be 
added so that the correct symmetry is realized for identical
particles. 
Disconnected diagrams have to be dropped. 
Only topologically different diagrams are allowed.  
With the self-energy that is obtained by opening one Green's function line, we 
can evaluate the density as done in this work.
A cluster decomposition for $\Phi$ is shown in Fig.\ \ref{Fig:1}. 

A general formalism in the chemical picture is not available at present. 
The full $A$-nucleon (cluster) propagator may be decomposed into a bound and 
scattering contribution (\ref{Apropagator}), given diagrammatically in 
Fig.~\ref{Fig:2}.
The concept is to consider the bound state part of the $A$- particle 
propagator on the same footing as the single-particle propagator. 
We illustrate this concept by considering special approximations that have 
been used in the present work.
The relevant diagrams are selected by construction.

\begin{figure}[ht] 
\centerline{\includegraphics[width=0.2\textwidth]{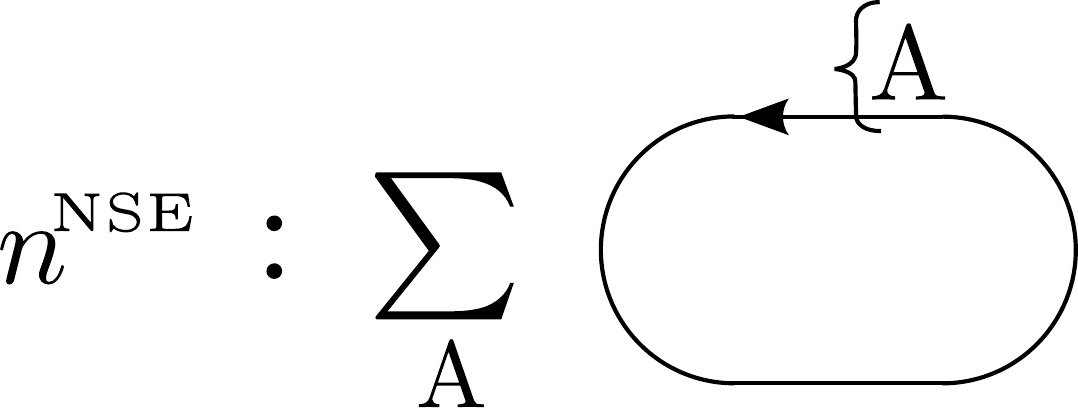}} 
\caption{Generating functional for the nuclear statistical equilibrium (NSE).} 
  \label{Fig:3} 
\end{figure}

The ideal gas of nucleons follows from the first diagram of Fig. \ref{Fig:1}. 
The Hartree-Fock (HF) contribution to the second virial coefficient follows 
from the second diagram and the corresponding exchange term. 
The sum of all ladder diagrams with $A=2$ yields a Bethe-Salpeter equation
that contains a scattering part and eventually a bound state part. 
For $A=2$, this binary approximation for $\Phi$ generates the standard 
Beth-Uhlenbeck formula.
In the chemical picture, we obtain the nuclear statistical equilibrium (NSE)  
supplementing the single nucleon propagator (first diagram of Fig.~\ref{Fig:1})
by the propagator of bound states, see Fig.~\ref{Fig:3}.
 
Within the single-nucleon approach, the corrections in lowest 
order of the interaction are given by the Hartree-Fock approximation.
\begin{figure}[ht] 
  \includegraphics[width=0.2\textwidth]{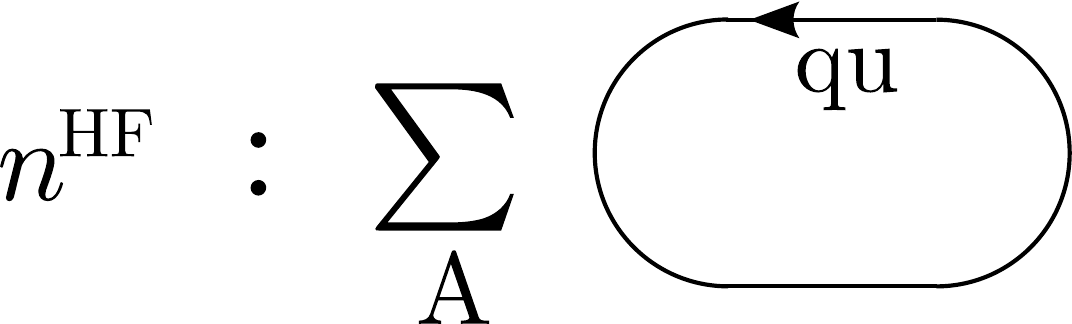}\\[5mm]
  \includegraphics[width=0.4\textwidth]{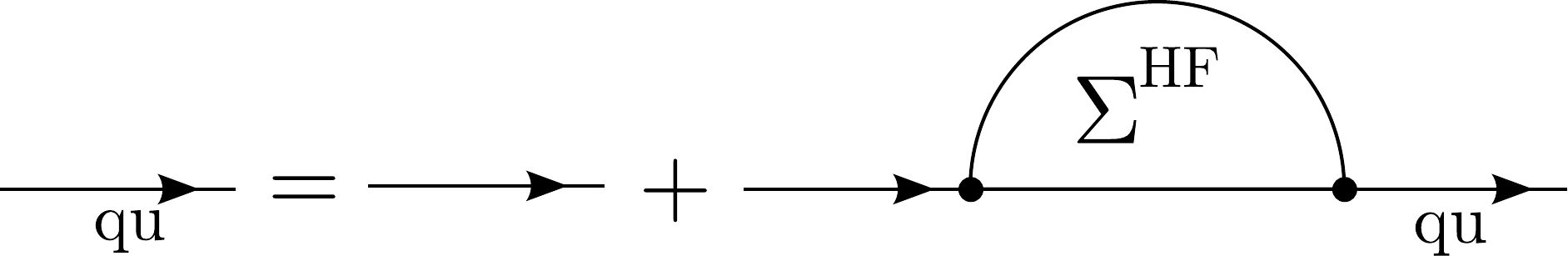}\\[5mm]
  \includegraphics[width=0.5\textwidth]{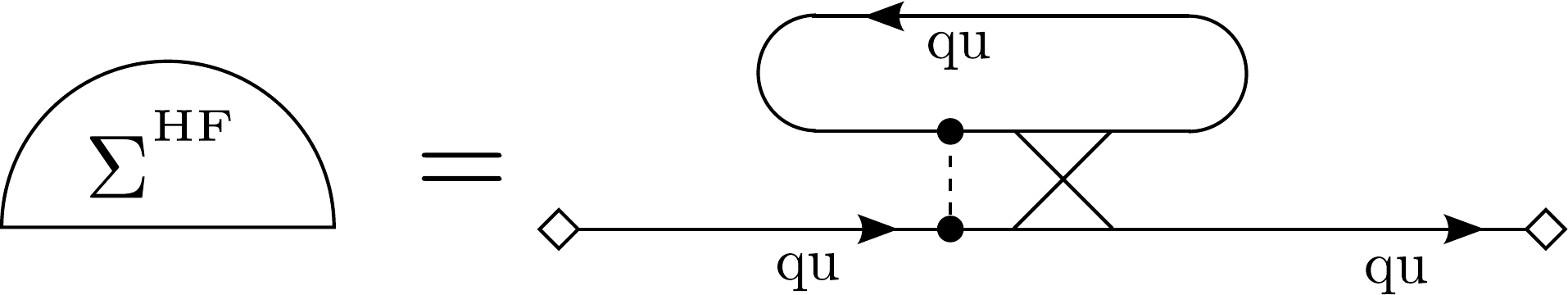}
\caption{ Hartree-Fock approximation for the generating functional
$\Phi^{\rm qu}$. 
The dotted line is the interaction. 
The crosses lines denote antisymmetrization. 
The diamonds at the endpoints of Green functions denote amputation 
(multiplication with the inverse).}
  \label{Fig:4} 
\end{figure}  
The corresponding Hartree-Fock approximation $\Phi^{\rm HF}$ (as the simplest 
version of a quasiparticle approximation) is shown in Fig.~\ref{Fig:4}.
The dotted line represents the interaction. 
The crossed lines denote antisymmetrization. 
Note that bookkeeping has to be respected.  
Diagrams that are included in  $\Phi^{\rm HF}$ have to be subtracted from 
other groups. 
For instance, the first order diagrams of the ladder sum in Fig.~\ref{Fig:1} 
(the second and third diagram) have to be subtracted.

\begin{figure}[ht] 
  \includegraphics[width=0.2\textwidth]{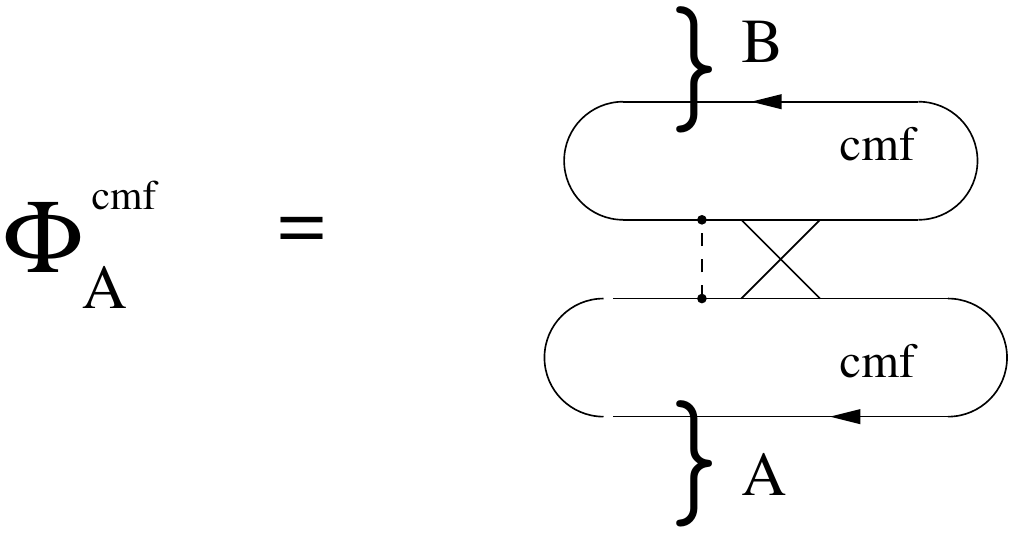}\\ [5mm]
  \includegraphics[width=0.45\textwidth]{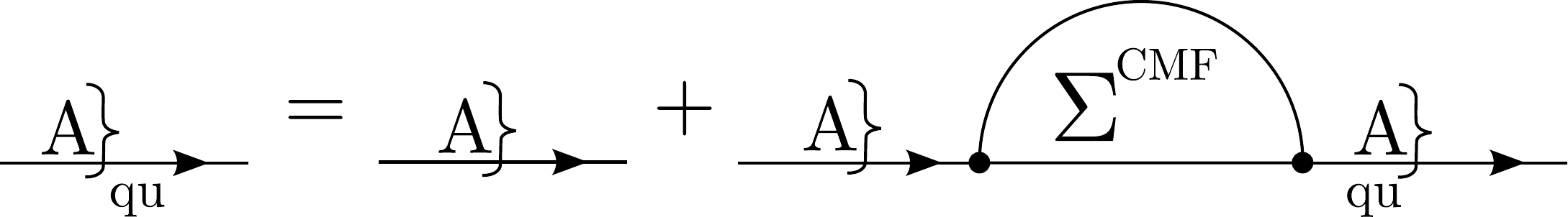}\\[5mm]
  \includegraphics[width=0.5\textwidth]{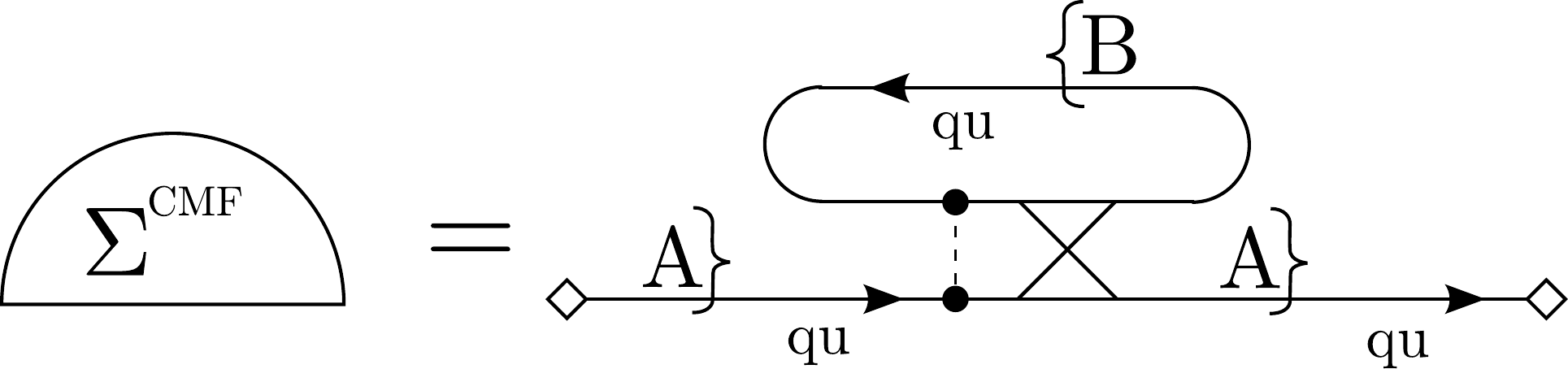}
  \caption{Quasiparticle propagator for cluster with cluster 
    mean-field self-energy. The crossed lines means the full 
    antisymmetrization between the clusters $A$ and $B$. 
 The diamonds at the endpoints of Green functions denote amputation 
(multiplication with the inverse).
}
  \label{Fig:7} 
\end{figure}

Turning to larger clusters, we can also define a quasiparticle 
propagator according to Eq.\ (\ref{bilinear}). 
The presentation by diagrams using the $A$-cluster self-energy is given in 
Fig.~\ref{Fig:7}. 
Then, $\Phi^{\rm HF}$ is completed according to the chemical picture by
the cluster mean-field approximation for the generating function 
$\Phi^{\rm CMF}$  shown in Fig.\ \ref{Fig:7}.
This approximation contains the cluster mean-field as described in 
App.~\ref{App:CMF}.
  
\end{document}